# How Fast is Your Detector? The Effect of Temporal Response on Image Quality

Tiarnan Mullarkey[1,2], Matthew Geever[1], Jonathan J. P. Peters[1], Ian Griffiths[3], Peter D. Nellist[3], Lewys Jones[1,4]

1. School of Physics, Trinity College Dublin, Dublin, Ireland.
2. Centre for Doctoral Training in the Advanced Characterisation of Materials, AMBER Centre, Dublin, Ireland.
3. Department of Materials, University of Oxford, Parks Road, Oxford, United Kingdom
4. Advanced Microscopy Laboratory, Centre for Research on Adaptive Nanostructures & Nanodevices (CRANN), Dublin, Ireland.

**Abstract**

With increasing interest in high-speed imaging should come an increased interest in the response times of our scanning transmission electron microscope (STEM) detectors. Previous works have previously highlighted and contrasted performance of various detectors for quantitative compositional or structural studies, but here we shift the focus to detector temporal response, and the effect this has on captured images. The rise and decay times of eight detectors' single electron response are reported, as well as measurements of their flatness, roundness, smoothness, and ellipticity. We develop and apply a methodology for incorporating the temporal detector response into simulations, showing that a loss of resolution is apparent in both the images and their Fourier transforms. We conclude that the solid-state detector outperforms the photomultiplier-tube (PMT) based detectors in all areas bar a slightly less elliptical central hole and is likely the best detector to use for the majority of applications. However, using tools introduced here we encourage users to effectively evaluate what detector is most suitable for their experimental needs.

## Introduction

The STEM has become a key piece of infrastructure in the toolkit of the modern researcher, ranging from biologists to material scientists (Pennycook & Nellist, 2011). This is perhaps largely due to its ability to produce readily interpretable images across a wide range of samples at varying magnifications using a finely focused electron beam (Dwyer, 2021; Fatermans et al., 2019). The necessity of using high energy electrons to achieve high resolution comes with the unavoidable consequence of associated damage to the samples being imaged. Avoiding damage is of great importance for imaging samples in their true state, or sometimes even at all, as losing a sample is regrettably common when imaging particularly fragile samples such as zeolites or biological specimens (Thach & Thach, 1971; Treacy & Newsam, 1987; Sartori Blanc et al., 1998; Egerton, 2019). An increasing understanding of these issues has also led to enthusiastic development of sample damage mitigation strategies such as cryogenic holders, and particularly a move away from long, slow scan-frames towards higher-speed, multi-frame data sets (Jones et al., 2018; Ophus et al., 2016; Jones et al., 2015; Mullarkey et al., 2022).

This ever-growing interest in low-dose (here measured in $e^-/Å^2$) STEM imaging has led to fast-scanning multi-frame acquisition strategies becoming increasingly commonplace. While lowering the beam current decreases the electron-dose, it is by fast-scanning that we can also decrease the electron *dose-rate*, which can be the more relevant factor than just total-dose for many materials (Johnston-Peck et al., 2016; Jiang & Spence, 2012; Egerton, 2019; Jones et al., 2018). Although shorter dwell-times inherently lead to lower signal-to-noise ratios due to less signal collection, what may be less obvious is that previously largely unreported artefacts also begin to appear. The most apparent of these is signal streaking, which becomes visible when the decay time of the scintillator from single-electron impact events is greater than or of a similar duration to the dwell-time being used (Buban et al., 2010; Mullarkey et al., 2020; Jones & Downing, 2018). Hence, as low-dose and dose-rate imaging becomes more important, so too does the high-speed performance characteristics of the detector being used.

Previously with an emphasis on structural studies we have stressed the importance of the flatness of the detector used (Macarthur et al., 2014). Now, with an interest in speed we seek to revisit and update that earlier work for the community in the context of temporal response, while also including a new solid-state detector. The aims then of this paper are to explore the origin and effects of this signal streaking and to compare the performance of a range of detectors available today. By doing so we hope to inform both instrument users, purchasers, and detector manufacturers of the significance of this effect, and the best conditions to operate an instrument under to minimise it.

## Background

*Image Formation in the STEM*

To understand signal streaking, we first discuss image formation in the STEM. While TEM imaging uses broad-beam illumination of the sample to form an image directly on a CCD camera or photographic plate, STEM imaging uses a finely focused electron probe scanned in a raster fashion. The electron probe spends a fixed amount of time at each pixel (the pixel dwell-time) collecting signals of interest using various detectors (e.g., annular dark-field (ADF), bright-field (BF), and energy dispersive X-ray spectroscopy (EDS) detectors). In the specific case of annular dark-field detectors, the focus of this work, electrons are collected which have been scattered to a range of angles by the sample being studied. Electrons which are scattered to high angles have undergone Rutherford scattering by the sample's nuclei. Therefore, heavier nuclei or thicker sample regions result in more scattering, and in turn, brighter pixels. Images produced this way are commonly referred to as having Z-contrast or mass-thickness contrast. This is an incoherent, easily interpretable imaging mode, which has led to it being widely used by many (Nellist & Pennycook, 2000; Jones, 2016).

At every pixel the output of the detector is captured, returning the average value of the detector's output during the dwell-time, and producing the arbitrary pixel values we see in the image. Recently, pixelated STEM detectors have also been developed, but these operate orders of magnitude slower than scintillator-photomultiplier types (dwell-times of tens or hundreds of microseconds, as opposed to tens of nanoseconds) and hence are not the focus of this work. An important aspect of the beam rastering in STEM is that the image formed is *not* from a simultaneous acquisition mode, so each pixel has been captured at a different point in time. Hence, STEM images are particularly susceptible to some well-known time-varying sources of noise such sample-drift (whether mechanical or thermal), stray electromagnetic fields, and other environmental sources such as acoustic disturbances (von Harrach, 1995; Jones & Nellist, 2013).

A lesser-known time-dependant artefact arises due to the decay time of non-ideal detectors and the corresponding readout electronics. A common setup is to use a scintillator-PMT based detector with a scintillating crystal such as yttrium aluminium perovskite (YAP). In such a system, an electron which impinges on the scintillating crystal generates numerous photons which travel along a light-guide. These photons are then converted back to electrons via the photoelectric effect and are accelerated towards successive dynodes. At each dynode an accelerated electron has enough energy to liberate numerous other electrons upon collision, and in this way a single initial electron hitting the detector can produce millions of electrons, and therefore a readily detectable output signal (Sang & LeBeau, 2016; Ishikawa et al., 2014).

While YAP may have a decay time as low as 25 ns (Baccaro et al., 1995; Novák & Müllerová, 2009), when in combination with the photomultiplier and readout electronics it is not uncommon to see decay times greater than 1 µs associated with the event from a single electron impact (Mullarkey et al., 2020; Sang & LeBeau, 2016; Mittelberger et al., 2018). If this decay time is greater than, or of a similar duration to, the pixel dwell-time we begin to see the 'streaking' of signal between neighbouring pixels (**Figure 1**), which has also been observed in other literature (Buban et al., 2010). While dwell-times less than ~1 µs are not often used, newer scan generators can achieve dwell-times of 50 ns, e.g., point Electronic's DISS 6 and the Gatan DigiScan 3, and so signal streaking will become both more relevant, and severe, as these become more widespread.

This streaking results in a loss of high-resolution information in the image, which is also observable in the Fourier transform as a drop in intensity of the higher spatial frequency Fourier components. This effect reduces the accuracy and precision of STEM images in which it is present, as signal which should be localised entirely to one pixel is instead assigned to neighbouring pixels. Furthermore, as this is a temporal effect it is independent of magnification and persists across images captured at varying length scales.

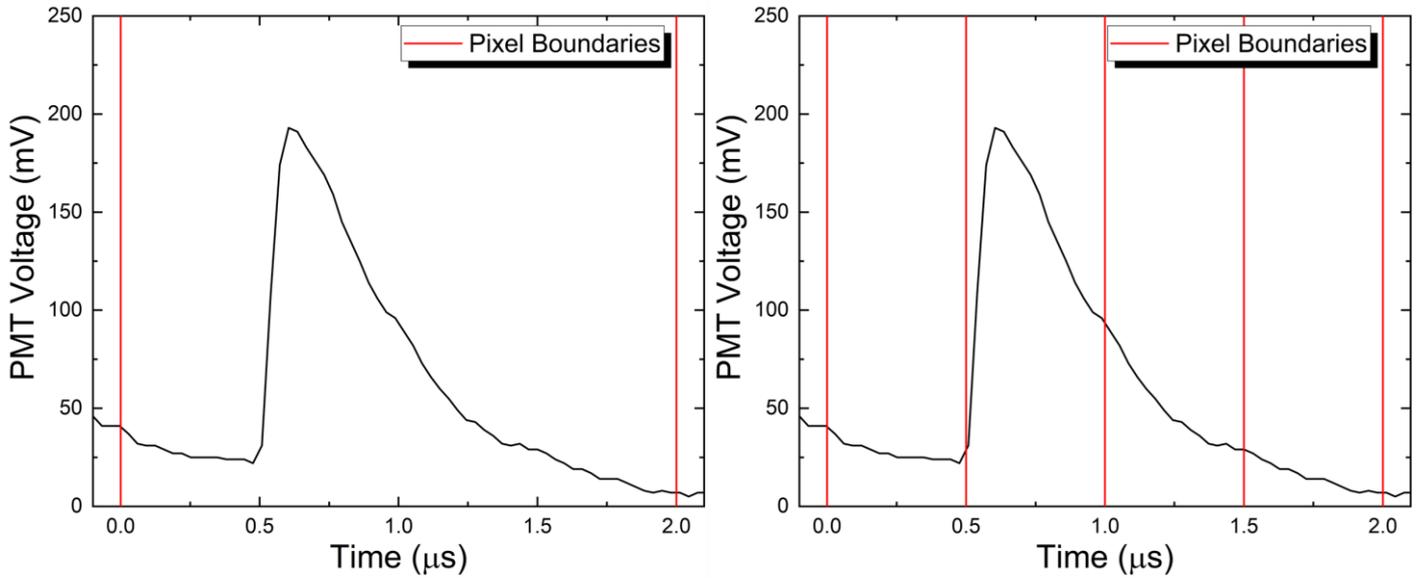

**Figure 1.** Signal streaking may occur at any dwell-time although it is rarely seen at dwell-times which are longer than the decay time of the pulse, such as on the left. However, it becomes increasingly common at shorter dwell-times, and eventually inescapable, such as on the right.

A further issue is that the response of the detector to electrons of the same energy is not homogenous. This is well known, and many detectors have had their responses quantified using metrics such as flatness, roundness, smoothness, and ellipticity (Macarthur et al., 2014). Some of these intensity variations in the response are inherent due to the asymmetric position of the readout electronics with respect to the detector or the presence of a hole through the centre of the scintillator. Others are because of issues which could be fixed at the manufacturing level, such as by combining the usual scintillating crystal with a layer of P47 scintillator to increase detection efficiency (Kaneko et al., 2014). Since the early days of STEM people discovered it was possible to do quantitative structural studies even with detectors which have large intensity variations (Singhal et al., 1997). Despite these known inhomogeneities, these detector imperfections persist through to other studies, with these detectors still being installed on machines to this day. To compensate for these effects, an operator will often directly capture an image of the detector (referred to as a detector scan or map) just before or after an experiment at the same conditions used for capturing the data. The detector response at these conditions is then used to normalise the data after subtraction of the dark level, allowing for easier comparison with simulations (Grieb et al., 2012; Mehrtens et al., 2013). A recent review of the literature in this field is presented in the following book chapter (Jones, 2020).

Whilst there have been approaches to remedy signal streaking such as signal-shifting or the digitisation of the raw signal from the detector, we understand that these approaches are not available to all (Jones & Downing, 2018; Mullarkey et al., 2020). Instead of focusing on the methods of removing streaking from images, we shift the focus to quantifying the effects this streaking has on images and the performance of different detectors on the market.

# Methods

*Capturing Single Electron Signals*

Understanding the decay behaviour of a detector requires observing the detector's response to single electrons. The method to achieve a low enough dose at the detector to see this signal may vary depending on the configuration of the microscope i.e., one can lower the extraction on a STEM with a cold field-emission gun (FEG) easily, while a STEM equipped with a Schottky FEG cannot simply lower the extraction. Instruments with monochromators are also easily able to reduce beam current by reducing the energy range of electrons which can pass. In the case of ADF detectors, moving the electron beam to a thin or low-mass region of the sample to reduce the amount of scattering to the detector and reducing the camera length such that most scattered electrons instead pass through the central hole are other widely available options. Although these are not ideal efficient operating conditions for imaging, they are suitable for collecting a weak signal for testing.

As image formation involves processing the detector output signal at each pixel, the raw trace of a single electron pulse is not preserved in the final image. Accessing this instead requires the use of an oscilloscope or similar. In this work pulses shown have either been captured on a PicoScope 2000 USB-streaming oscilloscope or a Red Pitaya measurement board. Where the raw data is accessed also depends on the microscope configuration. In some cases, the detector can be connected to directly (using for example a BNC connector), and if a microscope is controlled by an external scan controller (i.e., point electronic DISS6 or Gatan DigiScan) it is possible intercept the signal close to the scan controller inputs. To accurately capture an electron pulse, it is sometimes necessary to adjust the brightness and gain to ensure no clipping, high or low, of the signal, as well as a suitably large signal to background ratio. Having captured a single pulse, it can then be analysed with respect to its rise times and decay times.

*Detector Scans*

While observing a pulse directly contains information on *how* the intensity response of a detector to a single electron varies, only by capturing detector scan can it be revealed *why*. The method for obtaining detector scans again varies with instrument, although necessarily starts with moving to a region of vacuum in the sample so the detector can be imaged directly. The beam must then be focused at the detector plane, which can be achieved by switching to diffractive or confocal mode while the detector is inserted. It is recommended to discuss with a technician for the instrument of interest if you wish to perform a detector scan, as incorrect setup may temporarily 'burn' the detector surface. These detector scans are analysed by calculating the flatness, roundness, smoothness, and ellipticity of each, as defined in Macarthur et al., (2014). These definitions are reproduced here for ease of reading:

- **Flatness**: Detector sensitivity with respect to scattering angle (radially) after averaging azimuthally.
- **Roundness**: A measure of the consistency of the detector sensitivity around the detector (azimuthally) after averaging radially.
- **Smoothness**: The full-width-quarter-maximum of the active region of the normalised histogram.
- **Ellipticity**: The percentage of the major over the minor diameters of the inner angle opening.

*Simulation*

To not only investigate the sources of image-streaking but to also simulate its effects, code was developed to reproduce electron pulses and artificially streak simulated images. This process starts by producing a simulation with known fractional scattering of the beam at each pixel, using software such as Prismatic 2.0 (Rangel DaCosta et al., 2021). For a known electron-dose this fractional scattering can be converted to a number of electron impacts per pixel. These impacts follow a Poisson distribution in time (Sang & LeBeau, 2016), and can have (for example) Gaussian distributed intensities to emulate detector inhomogeneity. The simulated pulse shape is then assigned at each generated time-stamp following these distributions. The average pixel intensity value is calculated, and an image with the effects of signal streaking is produced. By changing the starting simulation, simulated pulse shape, dwell-time, and pulse intensity distribution, images which account for detector performance can now be simulated and studied relative to a known ground truth.

## Results & Discussion

*Detector Scans*

Detector scans from eight detectors produced by five different manufacturers were captured: one bright-field detector, six PMT based annular dark-field detectors, and one solid-state segmented annular detector (**Figure 2**). A large variety in their responses can be seen qualitatively, and this is also evidenced in the intensity profiles which are captured from across the centre of each detector. In some we see an asymmetric response which may imply that this is being caused by the placement of the readout electronics. For example, we know the light-guide and PMT for detector A are located on the left-hand side as presented in the image. Correspondingly, we see a region of low sensitivity on the right-hand side of the central hole, as many photons produced here are unable to navigate around the hole and reach the PMT. Similar behaviour can be seen across many of the other detectors.

|  | Company 1 | Company 2 | | | Company 3 | | Company 4 | Company 5 |
|---|---|---|---|---|---|---|---|---|
|  | Detector A | Detector B | Detector C | Detector D | Detector E | Detector F | Detector G | Detector I |
| **Detector Map & Profile** | 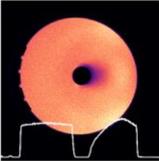 | 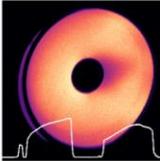 | 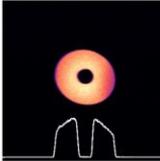 | 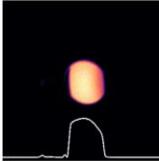 | 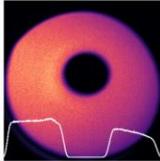 | 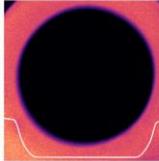 | 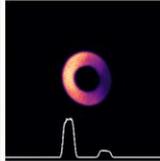 | 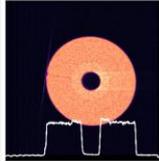 |
| **Flatness** | 9.86% | 6.69% | 9.61% | 8.97% | 8.38% | 4.07% | 21.60% | 2.94% |
| **Roundness** | 5.08% | 5.42% | 6.63% | 6.27% | 15.95% | 5.77% | 49.64% | 1.34% |
| **Smoothness** | 29.77% | 51.76% | 42.55% | 51.19% | 58.50% | 33.16% | 149.81% | 2.70% |
| **Ellipticity** | 1.88% | 12.37% | 2.81% | - | 4.90% | 1.88% | 10.69% | 2.82% |

**Figure 2.** Detector maps of the detectors used to capture individual pulses in **Figure 3** overlaid with their intensity profiles. Flatness, roundness, smoothness, and ellipticity measurements are tabulated below, with these values given as a measure of the deviation from a perfect detector, where 0% would be a perfect match. We choose not to show company names, but we encourage users to perform analysis themselves on their own system.

When comparing these detectors using the previously defined metrics, we see that no single detector performs best across every category. To first compare scintillator-photomultiplier tube-based detectors against each other we see that while detector A is the most round and smooth and tied for least elliptical, it is one of the least flat detectors. If compositional studies are important than this detector may be ill-advised due its poor flatness, despite its otherwise strong performance. For quantitative studies where comparisons with simulation will be used having a well-defined inner-angle is extremely important, and so detectors with large ellipticity values cannot be recommended. Note that ellipticity is undefined for the bright-field detector as it is defined relative to the inner angles in this study.

The solid-state detector outperforms the rest in every category except ellipticity. As electrons are converted directly to charge in this style of detector many of the problems and challenges associated with photomultiplier-based readouts and absent. However, there are defined channels through which the detector outputs are read, which result in very thin dead zones, perhaps impacting some detection efficiency. It can also be seen that the inner four rings of the segmented detector appear brighter than the outer rings, however such an offset could be eliminated with a gain reference at the beginning of an experiment.

*Electron Pulses*

We show here a comparison of individual pulses captured from some of the detectors presented above and discuss their 10-90% rise and decay times and other characteristics (**Figure 3**). We choose the rise and decay times to discuss as they can be intuitively related to important aspects of image degradation.

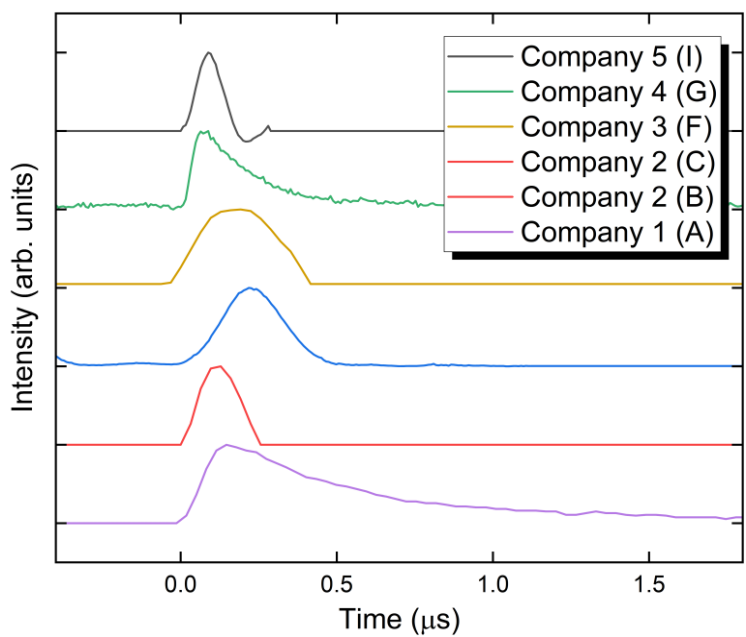

**Figure 3.** Example of single electron impacts from six different STEM ADF detectors spanning five different manufacturers, with the model specified by the letter in brackets. The intensity of a pulse varies due to detector inhomogeneity and gain, but each have been normalised here for comparative purposes. Some pulses have artificially flat backgrounds due to how the pulse profile was extracted from its respective data stream.

There are broadly two pulse shapes seen: a sharp rising edge followed by an exponential decay, and a more symmetric, Lorentzian/Gaussian pulse shape. While the former is perhaps more naturally understood as being caused by the electron-detector interaction, the latter may be due to shaping amplifiers, often used where preserving the area under the curve is important in applications such as spectrometers (Buzzetti et al., 2003).

In the case of the symmetric pulse-shape both rise-time and decay-time are important parameters when evaluating the likelihood, and severity, of pixel streaking. For the exponential decay shape, the decay time is often an order of magnitude larger than the rise time and is more significant when it comes to the severity of the streaking (**Figure 4**). The rise time however has another piece of significance. Should one look to digitisation approaches, having a sharp feature such as a very short rise time can aid in electron detection, and this can counteract the drawbacks caused by the decay time (Mullarkey et al., 2020).

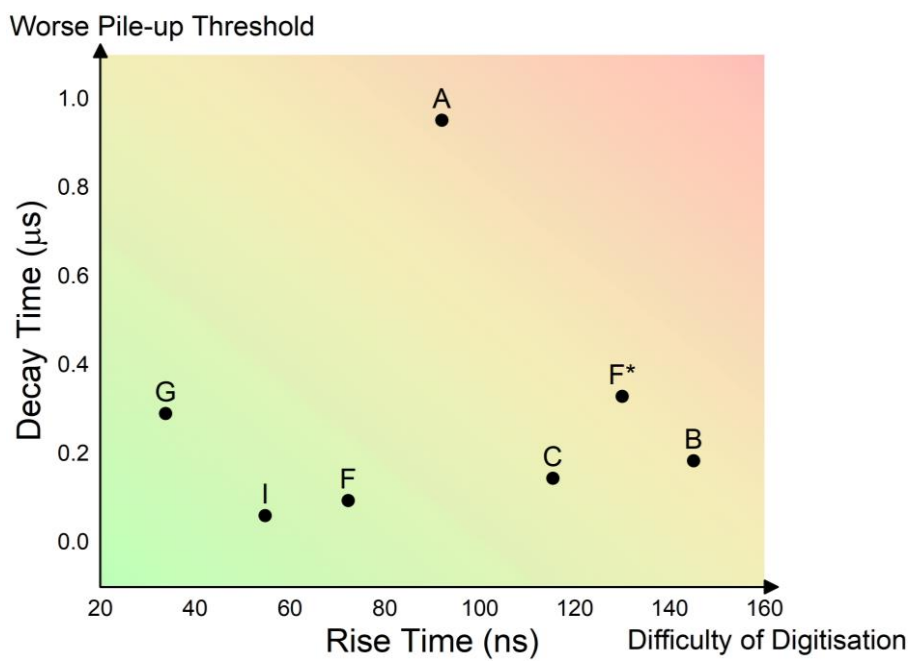

**Figure 4.** Typical values for the 10-90% rise times (in ns) and decay times (in µs) of pulses were calculated and are plotted here for the same detectors used to capture pulses in **Figure 3**. The closer a point is to the origin the less likely it is to streak, and the less severe the streaking will be if it occurs.

It was also noted that the shape of a pulse can depend on where it is captured from the instrument. This may be due to effects such as impedance matching and is seen here with two pulse sizes reported for detector F, with an example shown in Mullarkey et al., 2020. Although it may not be possible to take advantage of this fact, it is important to note for precise reporting of electron pulse sizes in literature.

*Simulation*

Having captured example electron pulses the first goal was to create a model of these. As an example, a lognormal distribution was used to recreate the pulse shape of detector A, shown in **Figure 5** alongside the experimental signal.

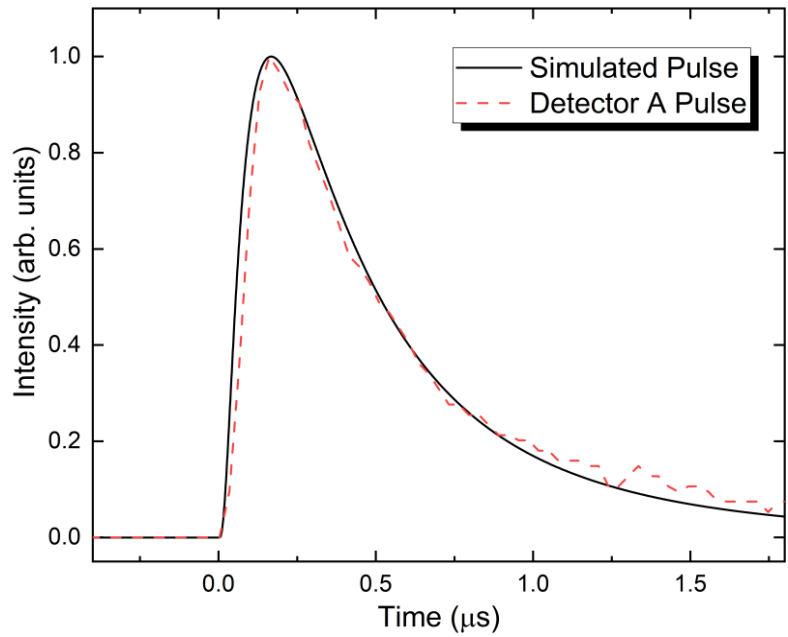

**Figure 5.** An example of an electron pulse captured from detector A (dashed, red), and our simulated model of it (solid, black).

This simulated pulse shape was then incorporated into a simulation using a dose of 5pA. The results of this are shown in **Figure 6** for simulations with dwell-times of 1 µs, 200 ns, 100 ns, and 50 ns, with this chosen as the shortest dwell-time as it corresponds to the speed attainable on new generation scan generators. As the simulated pulse has a decay time greater than 1 µs it is expected that streaking will be present in all images, but significantly more severe in the images with lower dwell-times. This behaviour is indeed seen in the images, with streaking occurring across multiple pixels being obvious in the 200 ns, 100 ns , and 50 ns dwell-time images. Although visually obvious, the Fourier transform can also be used as further evidence of image degradation. The Fourier transforms are presented in the lower row of the image, with one notable feature appearing as we move to shorter dwell-times: a drop in intensity of the higher spatial frequency Fourier components in the fast-scanning direction. This can be understood as the streaking resulting in a loss of high-frequency information in the images, reflected in this frequency envelope in the Fourier transform. It is worth reiterating that as this is a temporal effect, and not spatial, it does not only effect higher magnification images. Sharp image features (a few pixels in size in the fast-scan direction) regardless of physical size, can be obfuscated by this streaking, and valuable information lost.

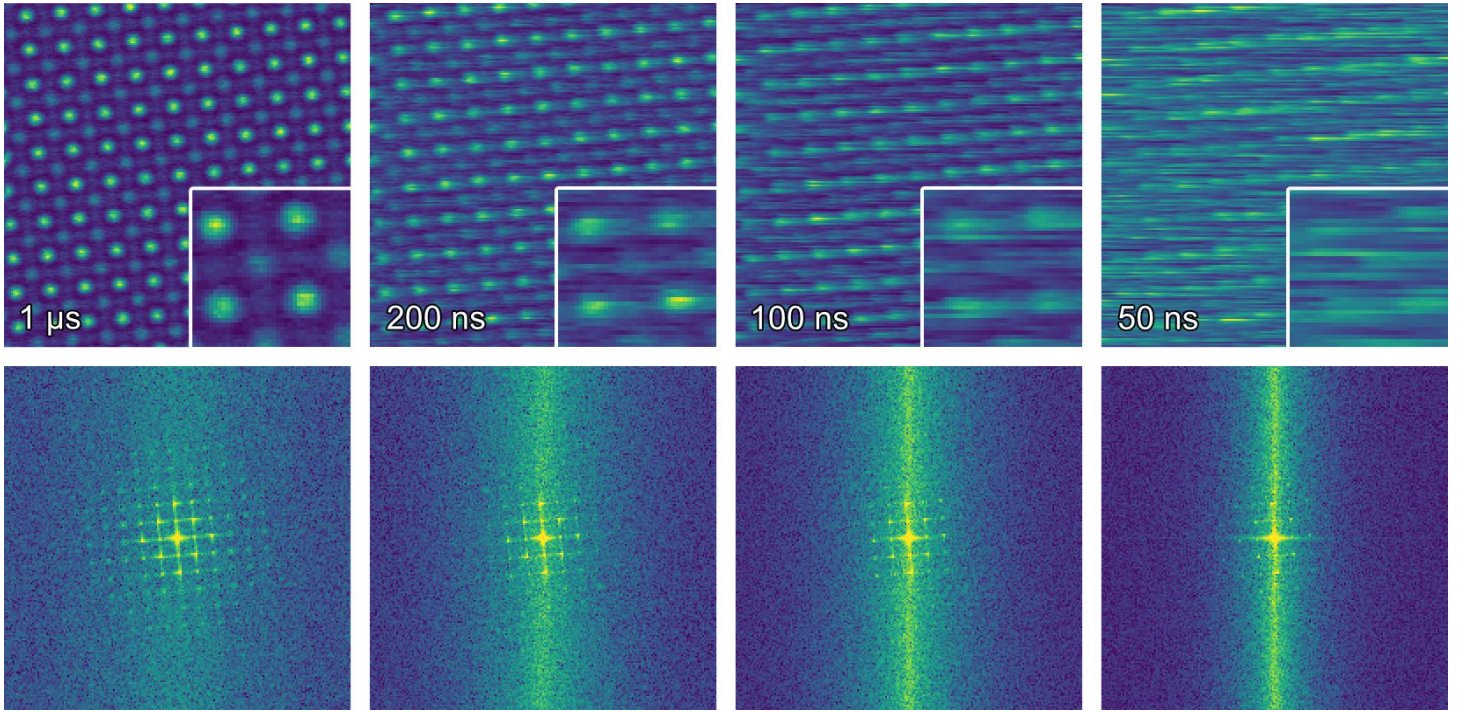

**Figure 6.** Four simulated streaked images of strontium titanate (SrTiO3) using a beam current of 5 pA are shown in the top row with dwell-times 1 µs, 200 ns, 100 ns, and 50 ns respectively. The pulse shape used to simulate the streaking is that shown in **Figure 5**. Below each image is its corresponding Fourier transform.

## Conclusions

In this paper we have shown single electron signals captured from various detectors and corresponding detector maps to highlight some of the issues caused directly by detector hardware, namely signal streaking and detector inhomogeneity. It is seen that pulses can vary between having sudden onsets and exponential decays, or symmetrical pulse shapes, perhaps due to the use of shaping amplifiers. The duration of these pulses varies both when comparing between technologies (PMT vs. solid-state), or even within the same class of detector. Rise times ranged from ~30 – 150 ns, and decay times from ~0.1- 0.9 µs for the pulses shown in this paper.

When comparing detector maps it was seen that no single detector outperformed in all metrics measured, although some do lead in certain categories. While detector A performs best in smoothness, roundness, and ellipticity and may appear a good choice, it also has a decay time more than twice as long as the other detectors, and so the worst high-speed performance. The solid-state detector outperforms in all categories but the ellipticity of its central hole (and even then, is only marginally worse), indicating that this is the nearest detector to an ideal case and that solid-state detection leads to a marked increase in performance. It is important to balance the contributions of the factors described in this paper when choosing a suitable detector for specific experimental uses.

Finally, by bringing these factors into simulations we have produced images containing streaking as it would appear on a detector A for a range of dwell-times. At shorter dwell-times we see obvious streaking between neighbouring pixels, and correspondingly a drop in the intensity at higher spatial frequencies in the Fourier transform. Even if streaking may not be visually obvious such as in the 1 µs dwell-time simulation, this drop in intensity is still present in the Fourier transform. By providing tools to properly characterise detector speeds, we hope to enable performance evaluation of detectors at the manufacturing level, or by prospective purchasers to gauge if the detector's performance will be suitable for their imaging needs. We also envision that as new scan generators which can scan an order of magnitude faster become more widely available, such evaluations will become increasingly more important should one want to produce images of acceptable quality at these now accessible speeds.

## Acknowledgments

The authors would like to acknowledge the Centre for Research on Adaptive Nanostructures and Nanodevices (CRANN) and the Advanced Materials and BioEngineering Research (AMBER) Network for financial and infrastructural support for this work. J.J.P.P. and L.J. acknowledge SFI grant 19/FFP/6813, T.M. acknowledges the SFI & EPSRC Centre for Doctoral Training in the Advanced Characterisation of Materials (award references 18/EPSRC-CDT-3581 and EP/S023259/1). This project has received funding from the European Union's Horizon 2020 research and innovation programme under grant agreement No 823717 – ESTEEM3.

**Competing Interests**: The authors declare no competing interests.

# References

BACCARO, S., BLAŽEK, K., DE NOTARISTEFANI, F., MALY, P., MARES, J., PANI, R., PELLEGRINI, R. & SOLURI, A. (1995). Scintillation properties of YAP:Ce. *Nuclear Instruments and Methods in Physics Research Section A: Accelerators, Spectrometers, Detectors and Associated Equipment* **361**, 209–215.

BUBAN, J. P., RAMASSE, Q., GIPSON, B., BROWNING, N. D. & STAHLBERG, H. (2010). High-resolution low-dose scanning transmission electron microscopy. *Journal of Electron Microscopy* **59**, 103–112.

BUZZETTI, S., GUAZZONI, C. & LONGONI, A. (2003). EROIC: a BiCMOS pseudo-gaussian shaping amplifier for high-resolution X-ray spectroscopy. *Nuclear Instruments and Methods in Physics Research Section A: Accelerators, Spectrometers, Detectors and Associated Equipment* **512**, 150–156.

DWYER, C. (2021). Quantitative annular dark-field imaging in the scanning transmission electron microscope—a review. *Journal of Physics: Materials* **4**, 042006.

EGERTON, R. F. (2019). Radiation damage to organic and inorganic specimens in the TEM. *Micron* **119**, 72–87.

FATERMANS, J., VAN AERT, S. & DEN DEKKER, A. J. (2019). The maximum a posteriori probability rule for atom column detection from HAADF STEM images. *Ultramicroscopy* **201**, 81–91.

GRIEB, T., MÜLLER, K., FRITZ, R., SCHOWALTER, M., NEUGEBOHRN, N., KNAUB, N., VOLZ, K. & ROSENAUER, A. (2012). Determination of the chemical composition of GaNAs using STEM HAADF imaging and STEM strain state analysis. *Ultramicroscopy* **117**, 15–23.

VON HARRACH, H. S. (1995). Instrumental factors in high-resolution FEG STEM. *Ultramicroscopy* **58**, 1–5.

ISHIKAWA, R., LUPINI, A. R., FINDLAY, S. D. & PENNYCOOK, S. J. (2014). Quantitative annular dark field electron microscopy using single electron signals. *Microscopy and Microanalysis* **20**, 99–110.

JIANG, N. & SPENCE, J. C. H. (2012). On the dose-rate threshold of beam damage in TEM. *Ultramicroscopy* **113**, 77–82.

JOHNSTON-PECK, A. C., DUCHENE, J. S., ROBERTS, A. D., WEI, W. D. & HERZING, A. A. (2016). Dose-rate-dependent damage of cerium dioxide in the scanning transmission electron microscope. *Ultramicroscopy* **170**, 1–9.

JONES, L. (2016). Quantitative ADF STEM: acquisition, analysis and interpretation. *IOP Conference Series: Materials Science and Engineering* **109**, 012008.

JONES, L. (2020). Practical Aspects of Quantitative and High-Fidelity STEM Data Recording. In *Scanning Transmission Electron Microscopy*, Bruma, A. (Ed.), pp. 1–40. CRC Press.

JONES, L. & DOWNING, C. (2018). The MTF & DQE of Annular Dark Field STEM: Implications for Low-dose Imaging and Compressed Sensing. *Microscopy and Microanalysis* **24**, 478–479.

JONES, L. & NELLIST, P. D. (2013). Identifying and correcting scan noise and drift in the scanning transmission electron microscope. *Microscopy and Microanalysis* **19**, 1050–1060.

JONES, L., VARAMBHIA, A., BEANLAND, R., KEPAPTSOGLOU, D., GRIFFITHS, I., ISHIZUKA, A., AZOUGH, F., FREER, R., ISHIZUKA, K., CHERNS, D., RAMASSE, Q. M., LOZANO-PEREZ, S. & NELLIST, P. D. (2018). Managing dose-, damage- and data-rates in multi-frame spectrum-imaging. *Microscopy* **67**, 98–113.

JONES, L., YANG, H., PENNYCOOK, T. J., MARSHALL, M. S. J., VAN AERT, S., BROWNING, N. D., CASTELL, M. R. & NELLIST, P. D. (2015). Smart Align—a new tool for robust non-rigid registration of scanning microscope data. *Advanced Structural and Chemical Imaging* **1**, 1–16.

KANEKO, T., SAITOW, A., FUJINO, T., OKUNISHI, E. & SAWADA, H. (2014). Development of a high-efficiency DF-STEM detector. *Journal of Physics: Conference Series* **522**, 012050.

MACARTHUR, K. E., JONES, L. B. & NELLIST, P. D. (2014). How flat is your detector? Non-uniform annular detector sensitivity in STEM quantification. *Journal of Physics: Conference Series* **522**, 1198–1199.

MEHRTENS, T., SCHOWALTER, M., TYTKO, D., CHOI, P., RAABE, D., HOFFMANN, L., JÖNEN, H., ROSSOW, U., HANGLEITER, A. & ROSENAUER, A. (2013). Measurement of the indium concentration in high indium content InGaN layers by scanning transmission electron microscopy and atom probe tomography. *Applied Physics Letters* **102**, 132112.

MITTELBERGER, A., KRAMBERGER, C. & MEYER, J. C. (2018). Software electron counting for low-dose scanning transmission electron microscopy. *Ultramicroscopy* **188**, 1–7.

MULLARKEY, T., DOWNING, C. & JONES, L. (2020). Development of a Practicable Digital Pulse Read-Out for Dark-Field STEM. *Microscopy and Microanalysis* **27**, 99–108.

MULLARKEY, T., PETERS, J. J. P., DOWNING, C. & JONES, L. (2022). Using Your Beam Efficiently: Reducing Electron Dose in the STEM via Flyback Compensation. *Microscopy and Microanalysis* **28**, 1428–1436.

NELLIST, P. D. & PENNYCOOK, S. J. (2000). The principles and interpretation of annular dark-field Z-contrast imaging. In *Advances in Imaging and Electron Physics* vol. 113, pp. 147–203.

NOVÁK, L. & MÜLLEROVÁ, I. (2009). Single electron response of the scintillator-light guide-photomultiplier detector. *Journal of Microscopy* **233**, 76–83.

OPHUS, C., CISTON, J. & NELSON, C. T. (2016). Correcting nonlinear drift distortion of scanning probe and scanning transmission electron microscopies from image pairs with orthogonal scan directions. *Ultramicroscopy* **162**, 1–9.

PENNYCOOK, S. J. & NELLIST, P. D. (2011). *Scanning Transmission Electron Microscopy*. Pennycook, Stephen J. & Nellist, Peter D. (Eds.). New York, NY: Springer New York.

RANGEL DACOSTA, L., BROWN, H. G., PELZ, P. M., RAKOWSKI, A., BARBER, N., O'DONOVAN, P., MCBEAN, P., JONES, L., CISTON, J., SCOTT, M. C. & OPHUS, C. (2021). Prismatic 2.0 – Simulation software for scanning and high resolution transmission electron microscopy (STEM and HRTEM). *Micron* **151**, 103141.

SANG, X. & LEBEAU, J. M. (2016). Characterizing the response of a scintillator-based detector to single electrons. *Ultramicroscopy* **161**, 3–9.

SARTORI BLANC, N., STUDER, D., RUHL, K. & DUBOCHET, J. (1998). Electron beam-induced changes in vitreous sections of biological samples. *Journal of Microscopy* **192**, 194–201.

SINGHAL, A., YANG, J. C. & GIBSON, J. M. (1997). STEM-based mass spectroscopy of supported Re clusters. *Ultramicroscopy* **67**, 191–206.

THACH, R. E. & THACH, S. S. (1971). Damage to Biological Samples Caused by the Electron Beam during Electron Microscopy. *Biophysical Journal* **11**, 204–210.

TREACY, M. M. J. & NEWSAM, J. M. (1987). Electron beam sensitivity of zeolite L. *Ultramicroscopy* **23**, 411–419.